\documentclass[12pt]{article}
\usepackage{amssymb}
\usepackage{amsmath,amsthm}
\usepackage{amsfonts}
\usepackage{graphicx}
\usepackage{appendix}
\usepackage{braket}
\usepackage{graphics}
\usepackage{indentfirst}

\usepackage[colorlinks=true
  ,urlcolor=blue
  ,anchorcolor=blue
  ,citecolor=cyan
  ,filecolor= magenta
  ,linkcolor= magenta
  ,menucolor= magenta
  ,linktocpage=true
  ,pdfproducer=medialab
  ,pdfa=true
]{hyperref}

\begin{document}

%%%%%%%%%%%%%%%%%%%%%%%%%%%%%%%%%%%
%%%%%%%%%%%%%%%%%%%%%%%%%%%%%%%%%%%
%%    THE TITLE PAGE    %%%%%%%%%%%
%%%%%%%%%%%%%%%%%%%%%%%%%%%%%%%%%%%
%%%%%%%%%%%%%%%%%%%%%%%%%%%%%%%%%%%
\

    \begin{center}
        {\huge\textbf{Dressed D-strings with
        Instability and Transverse Rotation:
        The Open String Pair Production
        }}\\
\vspace*{.95cm}
{\Large  \textsc{Hamidreza Daniali}\footnote{\textsc{Email:} 
\href{mailto:hrdl@aut.ac.ir}{\texttt{hrdl@aut.ac.ir}} } \&
\textsc{Davoud Kamani}\footnote{\textsc{Email:} \href{mailto:
kamani@aut.ac.ir}{\texttt{kamani@aut.ac.ir}}
(Corresponding author)}}\\

\vspace*{.5cm}
 \it{Department of Physics, Amirkabir University of
Technology (Tehran Polytechnic) \\
P.O.Box: 15875-4413, Tehran, Iran \\}
    \end{center}
\vspace*{1.25cm}
    \begin{abstract}

Motivated by the Schwinger effect in the QED,
we investigate the open string pair creation 
for the dressed D1-branes, incorporating 
electric and tachyonic fields and also transverse rotation
in the presence of the antisymmetric (Kalb-Ramond) 
background field. The background spacetime is 
partially compact on a torus. 
Our first observation is that the presence of the  
tachyonic field together with the 
transverse rotation prevents the open string 
production. Consequently, we shall quench the 
tachyonic field. Thus, we find that the pair creation 
arises only when the angular frequencies of the 
D-strings satisfy a rational relation. 
Besides, we observe that the compactification 
enhances the production rate of the open strings. 
Finally, we study various special cases of the system. 

\end{abstract}
    \ \\
\begin{flushleft}
\textsc{PACS numbers:} 11.25.-w; 11.25.Wx; 98.62.Dm\\
\vspace*{.25cm}
\textsc{Keywords:} Dressed D-string; Schwinger effect;
Open string; Pair production; Transverse rotation;
Brane instability.
\end{flushleft}

\newpage

%%%%%%%%%%%%%%%%%%%%%%%%%%%%%%%%%%%%%
%%%%%%%%%%%%%%%%%%%%%%%%%%%%%%%%%%%%%
%%%%%%%%    THE BODY    %%%%%%%%%%%%%
%%%%%%%%%%%%%%%%%%%%%%%%%%%%%%%%%%%%%
%%%%%%%%%%%%%%%%%%%%%%%%%%%%%%%%%%%%%
\section{Introduction}
\label{100}
%%%%%%%%%%%%%%%%%%%%%%%%%%%%%%%%%%%%%
%%%%%%%%%%%%%%%%%%%%%%%%%%%%%%%%%%%%%

The D-branes interactions are very important in
the superstring theories \cite{1}, \cite{2}, \cite{3}.
When two D-branes are
parallel to each other, for any given separation,
the lowest order stringy interactions can be applied,
which are open/closed string channels \cite{1}-\cite{5}.
The first formulation involves the one-loop diagram
of an open string, where the ends of the open string are
attached to the D-branes.
Alternatively, the D-branes interaction can
be represented by a closed string tree-level
diagram. In this approach, each D-brane is described via
a closed string boundary state. Hence,
one of the D-branes emits a closed string.
The emitted closed string propagates between the
D-branes, and then it is
absorbed by the second D-brane \cite{1}-\cite{11}.

A brane can be dressed with a $U(1)$ internal
gauge potential, and also it can have tangential
or transverse motion \cite{12}-\cite{19}.
For such branes the interaction gives an attractive force.
For the configurations with a large separation, the
attractive force is dominantly mediated by the exchange
of the massless closed string modes.
Conversely, in the small branes separations, the
nature of the interaction - attractive
or repulsive - becomes indeterminate.
Thus, the open string formulation
provides the most precise framework for
analyzing such cases. In these cases,
when the branes have been dressed with
the electric fluxes, the open string pairs
are produced. The electric fluxes on
the branes obviously provide the energy
of the open string pair creation,
similar to the Schwinger pair production.
The pair creation
implies that the interaction amplitude of the branes
includes an imaginary part \cite{20}, \cite{21}.
In fact, the rate of the open string pair production
is very small. When the magnetic fluxes are
put on the branes this rate is
extremely changed \cite{22}-\cite{27}.

The open string pair production has
been usually investigated for
the parallel or oblique configurations
of the branes \cite{22}-\cite{28}.
In this paper, we propose a setup in which
the branes undergo dynamical rotations about
one of their common transverse directions.
The branes have been dressed with different
$U(1)$ gauge potentials and tachyonic
fields. Besides, the background $B$-field
has filled the spacetime.
The background spacetime is partially compact
on a torus with various compactification radii.
To ensure the absence of intersection,
our investigation will be restricted to the
D1-branes (D-strings). Our computations will be in
the framework of the bosonic string theory.

It will be proved that for occurring the open
string pair production, the tachyonic fields
on the D-strings of our system must be quenched,
and also the rotational frequencies
of the D-strings should obey a specific rational 
relation. In other words, the open 
string pair creation does 
not occur via the interaction of the transversely rotating 
D-strings which carry both a tachyonic 
and $U(1)$ gauge potential. Besides, the D-strings cannot
intrinsically possess any magnetic fluxes
to enhance the pair production rate. However,
it will be shown that the compactification of the
spacetime enhances this rate.
This result is consistent with the literature
\cite{1}, \cite{2}, \cite{29}.

The rest of this paper is organized
as follows. In Sec. \ref{200}, we obtain the interaction
amplitude of the rotating D-strings in the
presence of all internal and background fields.
In Sec. \ref{300} the rate of the open string
pair creation is calculated. In Sec. \ref{400},
we analyze this rate for some special cases.
Sec. \ref{500} is devoted to the conclusions.
In the appendix. \ref{app} we compute the corresponding
boundary state of a dressed-rotating D-string.

%%%%%%%%%%%%%%%%%%%%%%%%%%%%%%%%%%%%%
%%%%%%%%%%%%%%%%%%%%%%%%%%%%%%%%%%%%%
\section{Feasibility of open string pair production}
\label{200}
%%%%%%%%%%%%%%%%%%%%%%%%%%%%%%%%%%%%%
%%%%%%%%%%%%%%%%%%%%%%%%%%%%%%%%%%%%%

In order to calculate the production of
the open string pairs,
at first we should compute the interaction amplitude.
Our system consists of two unstable
rotating D1-branes (D-strings) at a separation with the
angular frequencies $\omega_1$ and $\omega_2$.
They live in the partially compact spacetime
$\mathbf{T}^n \otimes \mathbb{R}^{1,d-n-1}$,
in the presence of a constant background $B$-field.
Besides, they have
been dressed with different $U(1)$ gauge potentials.
The signature of the spacetime metric is
$g_{\mu\nu}={\rm diag} (-1,1,\cdots, 1)$.

The interaction amplitude is
computed in the closed string channel,
which contains the exchange of the closed
strings between the interacting D-strings
\begin{eqnarray}
\mathcal A^{\rm CL}(t) = 2 \alpha'
\int_0^\infty {\rm d}\mathcal T \ ^{\rm (tot)}_{(t)}
\langle \mathbb B_1|
e^{-\mathcal T\mathcal H}
|\mathbb B_2\rangle^{\rm (tot)}_{(t)},
\label{1}
\end{eqnarray}
where $\mathcal H$ is the total Hamiltonian
of the closed string.

In this formalism, ``$t$'' denotes 
the temporal spacetime coordinate, 
specifically the eigenvalue of the 
zero-mode operator $x^0$ acting on 
the boundary state (see Appendix \ref{app}, 
especially after Eq. \eqref{A3}). 
This ensures that the rotation is uniform in the spacetime, 
and prevents the position-dependent 
of the angular velocities. 
In contrast, the integration variable 
``$\mathcal T$'' represents the 
proper time (modular parameter) on the 
closed-string worldsheet, which exchanges
between the D1-branes. Besides, the states
\begin{eqnarray}
    | \mathbb B_{1,2}\rangle_{(t)}^{\rm (tot)} =
    | \mathbb B_{1,2}\rangle_{(t)}^{(0)}
    \otimes | \mathbb B_{1,2}\rangle_{(t)}^{\rm (osc)}
    \otimes | \mathbb B\rangle^{\rm (g)},
    \end{eqnarray}
define the total boundary states, corresponding to the
D-branes. The amplitude \eqref{1} manifestly
represents the dynamical evolution of the
D-strings configuration.

According to the Appendix~\ref{app}, 
and employing the standard canonical 
commutation relations of quantum mechanics 
together with the integral representation of the 
Dirac delta function, as well as the detailed 
calculations in 
Refs. \cite{11}--\cite{15}, we obtain the following 
interaction amplitude for the D-strings
\begin{eqnarray}
    \mathcal A^{\rm CL} (t)
&=& \frac{\alpha'T^2_p}{2\sqrt{U'_1 U'_2}} \frac{
    \prod_{m=1}^\infty \left[\Phi_{(m)1}^\dagger (t)
    \Phi_{(m)2}(t)\right]^{-1}}{|\sin \Omega t|}
    \int_0^\infty {\rm d}\mathcal T \;
    \left(\sqrt{\frac{\pi}{\alpha'\mathcal T}}
        \right)^{{\bar d}_n}e^{(d-2)\mathcal T/6}
\nonumber\\
    &\times& \sqrt{\frac{\pi}{\alpha'}
\left(\mathcal T-i\frac{U_1' - U'_2}{U_1'U_2'}\right)^{-1}}
    \;\exp\left( -\frac{\sum_{\bar i_n}(y_1^{\bar i_n}
    - y_2^{\bar i_n})^2}{4\alpha' \mathcal T}
    \right) \prod_{\bar i_c}
\Theta_3 \left(\frac{y_1^{\bar i_c}
    - y_2^{\bar i_c}}{2\pi R_{\bar i_c}} \right|\left.
    \frac{i\alpha' \mathcal T}{\pi R^2_{\bar i_c}}\right)
\nonumber\\
    &\times& \sum_{N_2} \exp\left\{
    \frac{N_2 R_2\left(\mathcal E_1 U'_2
    + \mathcal E_2 U'_1\right)}{\alpha'
\left(U'_1 U'_2\;\mathcal T+ i
    (U_2'- U_1') \right)}\right\} \prod_{n=1}^\infty
    \frac{(1-q^{2n})^{4-d}}
{\det({\bf 1}- D'_{(n)1} D_{(n)2}q^{2n})}\;\;,
\label{3}
\end{eqnarray}
where $q\equiv e^{-2\mathcal T}$, and
\begin{eqnarray}
D_{(n)} =
    \begin{pmatrix}
        \mathbf A_{(n)} & \mathbf C_{(n)} \\
        -\mathbf C_{(n)} & \mathbf B_{(n)}
    \end{pmatrix},\;\;\;\;\;\;\;\;
D'_{(n) IJ} = D^\dagger_{(n)IJ}\;,\;\;{I,J\in\{0,1\}}.
\end{eqnarray}
The sets $\{\bar i_n\}$ and $\{\bar i_c\}$
are the non-compact and
compact parts of the set $\{\bar i\}$, respectively,
and ${{\bar d}_n}$ (${{\bar d}_c}$) denotes the dimension of
the subspace $\{x^{\bar i_n}\}$ ($\{x^{\bar i_c}\}$).
Besides, $\Omega \equiv \omega_1 - \omega_2$
represents the relative frequency of the D-strings. 
The appearance of $|\sin \Omega t|$ in the denominator 
follows directly from the identity 
$\delta(a x) = |a|^{-1}\delta(x)$. 
This indicates that the amplitude in 
Eq. \eqref{3} is well-defined only for 
$\omega_1 \neq \omega_2 $. 
Consequently, the parallel configuration 
constitutes a distinct case and cannot be 
naively recovered by taking the limit 
$\Omega \to 0$.\footnote{For example see Refs. 
\cite{12, 14}, where analogous results have been 
derived for the moving D$p$-branes.}
In addition, we see the rotating system imposed
a time-dependent prefactor for the amplitude.
Thus, the strength of the interaction is periodic,
as expected. 

The effects of the toroidal compactification
in the amplitude \eqref{3} have been
represented by the Jacobi $\Theta_3$-function
and the summation. For reducing the setup to
the non-compact scenario, one should replace
$\Theta_3 \rightarrow 1$, $\sum_{N_2} \rightarrow 1$
$\{ \bar i_n\} \rightarrow \{\bar i\}$ and
$\{\bar i_c\} \rightarrow \varnothing$. 

In the computation of the production 
rate of the open string pairs, an essential 
step is to deform the determinant
in Eq. \eqref{3} to the feature
$(1-\Gamma q^{2n})(1-\Gamma' q^{2n})$
such that $\Gamma\Gamma'=1$. 
However, the existence of the tachyonic 
field, accompanied by the transverse rotation,
prevents these relations, and hence 
there is no open string pair creation.
Thus, for receiving a nonzero 
pair creation rate, the configuration should  
include either the bare-static D-strings, 
which gives the traditional results, 
or the tachyonic profile and field 
strength should be mode-dependent, 
which contradicts our initial assumption 
of the path-integrally solvable backgrounds. 
Therefore, through the parameter analysis, the 
open-string pair production requires stable 
(tachyon-free) D-strings. Besides, the following 
rational relation for the rotational frequencies is
necessary to obtain a non-trivial $SL(2,\mathbb{C})$ matrix
\begin{eqnarray}
\label{5}
\frac{\omega_1}{\omega_2} =
\frac{2n_1\pm 1}{2n_2\pm 1} \equiv {\mathcal K}
\;\;,\;\;\;{n_1,n_2 \in \mathbb Z}.
\end{eqnarray}
This demonstrates that the ratio ${\omega_1}/{\omega_2}$
must be rational.
If $\mathcal K > 0 $\footnote{Since in the denominator
of Eq. \eqref{3} we have $|\sin \Omega t|$, the case
$\mathcal K = 1$ is excluded.},
both D-strings rotate in the same direction
(either clockwise or counter-clockwise),
and for $\mathcal K < 0 $ they rotate in
the opposite directions. In the 
case of quantized frequencies, 
$\mathcal K$ must necessarily be odd. 
Consequently, the open string pair production is 
constrained by pulsing feature.

%%%%%%%%%%%%%%%%%%%%%%%%%%%%%%%%%%%%%
%%%%%%%%%%%%%%%%%%%%%%%%%%%%%%%%%%%%%
\section{The rate of the pair production}
\label{300}
%%%%%%%%%%%%%%%%%%%%%%%%%%%%%%%%%%%%%
%%%%%%%%%%%%%%%%%%%%%%%%%%%%%%%%%%%%%
In this section, we derive the
production rate of the open string pairs.
Given the specified conditions in Sec. \ref{200},
the interaction amplitude must reduce to
\begin{eqnarray}
\mathcal A_{\rm red}^{\rm CL} (t)&=& \frac{\alpha'
T_1^2 \ell}{2} \frac{\sqrt{(1-\mathcal E_1 \cos\omega_1 t)
(1-\mathcal E_2 \cos\omega_2 t)}}
{|\sin \Omega t|} \int_0^\infty d\mathcal T
\left(\sqrt{\frac{\pi}{\alpha' \mathcal T }}
\right)^{{\bar d}_n}  e^{(d-2){\mathcal T}/6 }
\nonumber\\
&\times &\exp\left( -\frac{\sum_{\bar i_n}(y_1^{\bar i_n}
- y_2^{\bar i_n})^2}{4\alpha' \mathcal T }
\right) \prod_{n=1}^\infty
\frac{(1-q^{2n})^{4-d}}{\det
\left(1- D'_{1} D_{2}q^{2n}\right)}
\nonumber\\
&\times& \prod_{\bar i_c} \Theta_3 
\left(\frac{y_1^{\bar i_c}
- y_2^{\bar i_c}}{2\pi R_{\bar i_c}} \right|\left.
\frac{i\alpha' \mathcal T }{\pi 
R^2_{\bar i_c}}\right) \Theta_3
\Bigg(0 \Bigg|\left. \frac{i \mathcal T  
R_1^2}{\pi \alpha'}
\left(\mathcal E_1^2 + \mathcal E_2^2 - \mathcal E_1
\mathcal E_2-1\right)\right),
\end{eqnarray}
where $\ell = 2\pi \delta(0)$ is the worldline length
of the D-string. The
mode-independent matrices $D$ and $D'$ are given by
$\{D,D'\} = \{D_{(n)}, D'_{(n)}\}\big|_{U'=0}$.
As it was mentioned, the determinant 
factor can be expressed as
$1- (\Gamma + \Gamma') q^{2n} + q^{4n}$.
Regardless of the explicit form of $\Gamma$
and $\Gamma'$ in terms of
the internal fields and frequencies, one can define
$\Gamma \equiv e^{2i\nu(t)}$
which gives $\Gamma + \Gamma'= 2\cos\left[ 2\nu (t)\right]$.
By imposing the condition that $\Gamma' = \Gamma^{-1}$,
we obtain

\begin{eqnarray}
\cos [2 \nu (t)] = 1+ \frac{2\left[{\mathcal E}_1 \cos
(\omega_1t)-{\mathcal E}_2\cos (\omega_2t)\right]^2
}{\left[1-{\mathcal E}_1^2 \cos^2 (\omega_1
t)\right]\left[1-{\mathcal E}_2^2 
\cos^2 (\omega_2 t)\right]} > 1 .
\label{7}
\end{eqnarray}
Hence, $\forall t$ we receive $\nu(t)$ is pure imaginary
$\nu(t) = i \tilde{\nu}(t)$.

By using the Jacobi $\Theta$- and Dedekind $\eta$-functions,
the amplitude takes the following feature
\begin{eqnarray}
\mathcal A_{\rm red}^{\rm CL} (t) &=& i\alpha' T_1^2
\ell \sinh \tilde \nu (t) \frac{\sqrt{(1-\mathcal E_1
\cos\omega_1 t)(1-\mathcal E_2 \cos\omega_2 t)}
}{|\sin \Omega t|} \int_0^\infty d \mathcal
T \ e^{(d-2){\mathcal T}/12}
\nonumber\\
    &\times & \left(\sqrt{\frac{\pi}{\alpha'
    \mathcal T}}\right)^{{\bar d}_n}
    \exp\left( -\frac{\sum_{\bar i_n}(y_1^{\bar i_n}
- y_2^{\bar i_n})^2}{4\alpha' \mathcal T} \right) \eta^{5-d}
    \left(\frac{i\mathcal T}{\pi}\right) \Theta_1^{-1}
    \left(\frac{i\tilde \nu (t) }{\pi}\Big|
    \frac{i\mathcal T}{\pi} \right)
\nonumber\\
&\times& \prod_{\bar i_c}\Theta_3 \left(\frac{y_1^{\bar i_c}
    - y_2^{\bar i_c}}{2\pi R_{\bar i_c}} \right|\left.
\frac{i\alpha' \mathcal T}
{\pi R^2_{\bar i_c}}\right)\Theta_3
    \Bigg(0 \Bigg|\left.
    \frac{i \mathcal T R_1^2}{\pi \alpha'}E\right),
\label{8}
\end{eqnarray}
in which ${\tilde \nu}(t)=-i{\nu}(t)$ 
was given by Eq. \eqref{7}, and $E$ is defined as
\begin{eqnarray}
E = \mathcal E_1^2 + \mathcal E_2^2
- \mathcal E_1 \mathcal E_2-1.
\end{eqnarray}
Since the $\Theta_1$-function is pure imaginary,
the amplitude \eqref{8} is real.

For a sufficiently large separation of the D-strings, i.e.,
${\mathcal T} \rightarrow \infty$, up to the leading order
which belongs to the exchange of the massless
closed strings, we obtain
${\mathcal A}^{\rm CL}_{\rm red}
({\mathcal T}\rightarrow \infty) >0$.
This reveals an attractive force between the D-strings
with the large distance.
However, in the small $\mathcal T$-integration,
the attraction or repulsion of the interaction becomes
ambiguous. Consequently, we employ the Jacobi transformation
${\mathcal T}\rightarrow 1/ {\mathcal T}$
to switch to the open string channel. By utilizing
the following identities
\begin{subequations}
\begin{align}
\Theta_1(\hat{\nu} \mid \hat{\tau})
& =i \frac{e^{-i \pi \hat{v}^2 /
\hat{\tau}}}{\sqrt{-i \hat{\tau}}}
\Theta_1\left(\left.\frac{\hat{v}}{\hat{\tau}}
\right\rvert\,-\frac{1}{\hat{\tau}}\right), \\
\Theta_3(\hat{\nu} \mid \hat{\tau})
& =\frac{e^{-i \pi \hat{v}^2 /
\hat{\tau}}}{\sqrt{-i \hat{\tau}}}
\Theta_3\left(\left.\frac{\hat{v}}{\hat{\tau}}
\right\rvert\,-\frac{1}{\hat{\tau}}\right), \\
\eta(\hat{\tau}) & =\frac{1}{\sqrt{-i \hat{\tau}}}
\eta\left(-\frac{1}{\hat{\tau}}\right),
\end{align}
\end{subequations}
the amplitude finds the form
\begin{eqnarray}
\mathcal A_{\rm red}^{\rm OP} (t) &=& \frac{\pi \ell
T_1^2 R_1 \sinh \tilde\nu(t)}{
|\sin (\Omega t)| \prod_{{\bar i}_c}
R_{{\bar i}_c}} \sqrt{\frac{\alpha'^{d-2}}{E} 
\Big(1-\mathcal E_1 \cos \omega_1 t\Big)
\Big(1-\mathcal E_2 \cos \omega_2 t\Big)}
\nonumber\\
&\times& \int_{0}^\infty d{\mathcal T}\Bigg[
\left(\frac{\pi}{\mathcal T}\right)^{
{\bar d}_n+1 }
\exp{\left(\frac{(d-2)\pi^2}{12{\mathcal T}}
-\frac{\tilde\nu^2 (t){\mathcal T}}{\pi^2}
-\frac{(Y_{\rm N}^2-Y_{\rm C}^2){\mathcal T}}{4\pi^2\alpha'}
\right)}
\nonumber\\
&\times& \frac{\Theta_3\left(0\Big|
\frac{i{\mathcal T} \alpha'}{\pi^2ER_1^2}\right)}{\eta^{d-5}
    \left(\frac{i{\mathcal T}}{\pi}\right) \Theta_1
    \left(\frac{\tilde \nu (t){\mathcal T}}{\pi^2}\Big|
    \frac{i{\mathcal T}}{\pi}\right)}
    \prod_{\bar i_c} \Theta_3
\left(\frac{-i{\mathcal T}R_{\bar i_c}Y_C}
{2\pi^2 \alpha'}\Bigg|
\frac{i{\mathcal T}R_{\bar i_c}^2}{\pi\alpha'}
\right)\Bigg],
\label{11}
\end{eqnarray}
where the squared distances between the D-strings, in the
non-compact and compact subspaces, are defined as   
$Y^2_{{\rm N}({\rm C})}\equiv \sum_{\bar i_n(\bar i_c)}
\left(y_1^{\bar i_n(\bar i_c)}
- y_2^{\bar i_n(\bar i_c)}\right)^2$.

To determine the effective mass of the stretched open
string between the D-strings, one can always substitute
the explicit forms of
the $\Theta_{1,3}$- and 
$\eta$- functions into Eq. \eqref{11},
which yields 
\begin{eqnarray}
{\mathcal M}^2_{\rm eff} (t) = \frac{1}{2\pi^2}
\left( \tilde\nu^2 (t) + \frac{Y^2_{\rm N} 
- Y^2_{\rm C}}{4 \alpha'}
- \frac{d-2}{6} \pi^2 \right) .
\label{12}
\end{eqnarray}
This equation indicates that when the squared
distance of the D-strings in the
compact subspace possesses dominant values,
a tachyonic shift occurs. In this case, the integrand of
Eq. \eqref{11} for small separations,
i.e. for ${\mathcal T}\rightarrow \infty$, diverges.
This is due to the tachyonic instability. Consequently, the 
tachyon condensation induces a phase transition \cite{30}.
However, for computing the rate of the open
string pair production, we work in
the configurations in which 
$\mathcal{M}^2_{\rm eff} \geq 0$.

According to the properties of the 
$\Theta$-functions ({\it e.g.},
see Refs. \cite{6}, \cite{31}),
for small separation of the D-strings, all parts of the
integrand in Eq. \eqref{11} are positive, except the
``\textit{sine}''-factor in the $\Theta_1$-function. 
This factor takes the values in 
the interval $(-1,1)$. The existence 
of this term leads to an
infinite number of simple poles along the positive 
${\mathcal T}$-axis.
Consequently, the amplitude possesses imaginary terms, which 
potentially suggests a new physical process, 
analogous to the Casimir effect.

In the residue calculations \cite{23}-\cite{26}, \cite{32},
the corresponding poles are
${\mathcal T}_k = k \pi / {\tilde\nu(t)}$, where
$k$ is any positive integer number.
Using the Schwinger formula for the pair production rate 
\begin{equation}
\mathcal W (t)= -2 \mathcal S^{-1} \ {\rm Im}\
\mathcal A_{\rm red}^{\rm OP} (t),
\end{equation}
we find the following feature for the rate of the 
creation of the open string pairs
per unit worldsheet area ${\mathcal S}$,
\begin{eqnarray}
\mathcal W (t) &=& \frac{\pi^{1- {\bar d}_n}
\ell T_1^2 R_1 [{\tilde \nu}(t)]^{{\bar d}_n}\sinh
\tilde{\nu} (t) }{{\mathcal S}| \sin \Omega t| 
\prod_{{\bar i}_c}R_{{\bar i}_c} }
\sqrt{\frac{\alpha'^{d-2}}{E} 
\Big(1-\mathcal E_1\cos\omega_1 t\Big)
\Big(1-\mathcal E_2\cos\omega_2 t\Big)}
\nonumber\\
&\times& \sum_{k=1} \Bigg\{
\frac{(-1)^{k+1}}{k^{{\bar d}_n+1}\;\prod_{n=1}^\infty 
\left[ 1- \exp{\left(- \frac{2k\pi^2}{\tilde\nu (t)}
\right)}\right]^{d-2}}\;
\exp\left[{\frac{(d-2) \tilde \nu (t)}{12 k}
-\frac{2k\pi^2}{\tilde\nu (t)}\mathcal M^{2}_{\rm eff}}
\right]
\nonumber\\
&\times& \Theta_3\left(0\Big|\frac{ik\alpha'}{\tilde \nu (t)  
ER_1^2}\right) \prod_{\bar i_c}\Theta_3\left(\frac{-ik
R_{\bar i_c}Y_{\rm C}}{2\tilde \nu (t)\alpha'}
\Bigg|\frac{ik\pi 
R_{\bar i_c}^2}{\tilde \nu (t)\alpha'}\right)
\Bigg\}\ . \qquad
\label{14}
\end{eqnarray}
As it was previously mentioned, the rate is positive.
Besides the open string pair 
creation, this rate explains the decay of the system too.
It becomes further evident when one 
contemplates the small separation
of the D-strings, essentially, when 
they are almost coincident. As we see,
the rate obviously depends on the various 
parameters of the configuration, i.e.,
the electric fields, angular frequencies, 
compactification radii, and dimensions of the 
compact and non-compact subspaces of the spacetime.
Therefore, one can adjust these parameters 
to receive a desirable value for the 
production rate. 

The Eqs. \eqref{12} and \eqref{14} reveal that 
by increasing the separation between the D-strings in the 
non-compact directions, i.e. $Y_{\rm N}$,
the mass of each pair substantially 
grows, and hence, the pair creation rate decreases.
Consequently, increasing the mass specifies that 
the probability of generating 
the heavy open string pairs is low.
However, by increasing the separation $Y_{\rm C}$ and 
the spacetime dimension 
the probability of producing a pair of 
the light open strings is increased.

As we see, the production rate is a complicated
function of the electric fields.
A nontrivial rate implies that 
at least one of the electric 
fluxes must be nonzero. From the physical 
point of view, the presence of an electric 
flux is essential to polarize
the area between the D-strings. Otherwise,
the pair creation of the open 
strings does not take place.

%%%%%%%%%%%%%%%%%%%%%%%%%%%%%%%%%%%%%
%%%%%%%%%%%%%%%%%%%%%%%%%%%%%%%%%%%%%
\section{The pair production in the special cases}
\label{400}
%%%%%%%%%%%%%%%%%%%%%%%%%%%%%%%%%%%%%
%%%%%%%%%%%%%%%%%%%%%%%%%%%%%%%%%%%%%

By decompactifying the compact directions,
we receive the non-compact spacetime.
For obtaining the production rate in this 
spacetime the following conditions should be applied:
$\Theta_3 \rightarrow 1$, 
$\{ \bar i_n\} \rightarrow \{\bar i\}$,
and $\{\bar i_c\} \rightarrow \varnothing$. 
Therefore, we acquire 
\begin{eqnarray}
\mathcal W^{\rm NC} (t) &=& \frac{\pi\ell 
T_1^2 \sinh \tilde\nu(t)}{\mathcal S
|\sin \Omega t|} \sqrt{\frac{\tilde 
\nu (t)[1-{\mathcal E}_1 \cos(\omega_1 t)]
[1-{\mathcal E}_2 \cos(\omega_2 t)]}{\alpha'^{d-5}}}
\nonumber\\
&\times& \sum_{k=1}
\frac{(-1)^{k+1}}{k^{3/2}}\frac{\exp\left[\frac{(d-2) 
{\tilde \nu} (t)}{12 k}
-\frac{2k\pi^2}{{\tilde \nu}(t)}
{\mathcal M}'^2_{\rm eff}\right]}{\prod_{n=1}^\infty
\left[ 1- \exp{\left(- \frac{2k\pi^2}{\tilde\nu (t)}
\right)}\right]^{d-2}}, \qquad
\end{eqnarray}
where 
\begin{eqnarray}
\mathcal M'^2_{\rm eff} &=& \frac{1}{2\pi^2}
\left( \tilde\nu^2 (t) + \frac{\bar Y^2}{4 \alpha'}
-\frac{d-2}{6} \pi^2 \right),
\nonumber\\
{\bar Y}^2 &=& \sum_{\bar i}\left(y_1^{\bar i}
- y_2^{\bar i}\right)^2. 
\end{eqnarray}

According to the leading terms of the $\Theta_3$-functions
in the compact scenario, in comparison to the non-compact
case, the rate enhances if and only if
\begin{eqnarray}
R_1 & > & \sqrt{E}\;\left({\tilde \nu}_{\rm max}
\right)^{(1-2 {\bar d}_n)/2} \;
\prod_{{\bar i}_c}R_{{\bar i}_c}\;,
\nonumber\\
{\tilde \nu}_{\rm max} &=&\frac{1}{2}
\cosh^{-1}\left[1+ \frac{2\left({\mathcal E}_1 
-{\mathcal E}_2\right)^2
}{\left(1-{\mathcal E}_1^2 \right)\left(1-{\mathcal E}_2^2 
\right)}\right].
\label{17}
\end{eqnarray}
This elucidates that the decay of the 
system in the compact spacetime is faster 
than its decay in the non-compact case.

In the special case of ${\tilde \nu (t)} \rightarrow 0$, 
the electric fields become negligible. Thus,
we receive the bare D-strings. In this limit,
the pair production rate, in the 
non-compact spacetime scenario, finds the feature  
\begin{eqnarray}
{\mathcal W}^{\rm NC}|_{\tilde\nu (t) \rightarrow 0}
&\approx& \frac{\pi\ell T_1^2}{\mathcal S
|\sin \Omega t|} \sqrt{\frac{{\tilde \nu}^3 (t)
[1-\mathcal E_1\cos(\omega_1 t)]
[1-\mathcal E_2\cos(\omega_2 t)]}{{\alpha'^{d-5}}}}
\nonumber\\
&\times& \exp\left(-\frac{2k\pi^2}{\tilde \nu (t)}
{\mathcal M}'^2_{\rm eff}\right),
\end{eqnarray}
which is infinitesimal. In the compact spacetime,
the foregoing limit defines the production 
rate as in the following 
\begin{eqnarray}
{\mathcal W}|_{\tilde\nu (t)\rightarrow 0} &\approx&
\frac{\ell T_1^2 R_1 \left({\tilde\nu (t)}/{\pi}
\right)^{{\bar d}_n +1} }{\pi^2
{\mathcal S}|\sin\Omega t|
\prod_{{\bar i}_c}R_{{\bar i}_c} }
\sqrt{\alpha'^{d-2} E\;[1-\mathcal E_1 \cos(\omega_1 t)]
[1-\mathcal E_2 \cos(\omega_2 t)] }
\nonumber\\
&\times& \exp\left({-\frac{2 \pi^2}{\tilde\nu(t)}
{\mathcal M}_{\rm eff}^2}\right)
\Theta_3\left(0 \Big| \frac{i\alpha'\pi E}
{\tilde \nu(t) R_1^2}\right)
\prod_{\bar i_c} \Theta_3\left(\frac{ 
R_{\bar i_c}\;Y_{\rm C}}{2i \tilde\nu(t) \alpha'}
\Bigg|\frac{\pi i R_{\bar i_c}^2}
{\tilde\nu(t)\alpha'}\right). 
\label{19}
\end{eqnarray}
Depend on the values of the radii of compactification 
and the D-strings separation in the 
compact subspace, i.e. $Y_c$,
this rate may be vanished or becomes nonzero.
The non-vanishing case is a remarkable result. 
Unlike the non-compact case where the rate 
vanishes as $\tilde{\nu}(t)\!\to\!0$, the 
compactification sustains a non-zero rate.
Besides, it leaves the discrete KK/winding 
contributions \cite{33}, \cite{34}, \cite{35} 
through the $\Theta_3$ factors in Eq. \eqref{19}. 

%%%%%%%%%%%%%%%%%%%%%%%%%%%%%%%%%%%%%
%%%%%%%%%%%%%%%%%%%%%%%%%%%%%%%%%%%%%
\section{Conclusions}
\label{500}
%%%%%%%%%%%%%%%%%%%%%%%%%%%%%%%%%%%%%
%%%%%%%%%%%%%%%%%%%%%%%%%%%%%%%%%%%%%
By applying the boundary state formalism,
the interaction amplitude of two 
unstable D-strings, with electric fluxes
and transverse motions, was computed.
The background spacetime is partially 
compact on a  torus with difference radii.
Presence of the various parameters of the 
configuration dedicated a generalized 
form to the amplitude.

We demonstrated that the 
open string pair production does not occur when the 
D-strings simultaneously are unstable 
and undergo the rotations. 
Accordingly, we quench the tachyon fields. 
Besides, happening of the pair creation 
imposes that the relative angular frequencies 
must satisfy a rational relation, 
$\omega_1/\omega_2 \in \mathbb{Q}$.
This reveals that the two D-strings 
can rotate in the same or in opposite directions.

We obtained the production rate of the open strings.
The production rate obviously explains the decay 
of the system too. It depends on the various parameters 
of the configuration. Thus, it is possible to adjust
these parameters to obtain any desirable 
value for the production rate.
We observe that for the heavy open strings 
the production rate decreases. 
In the special limits where the rotational 
frequencies and/or the background fields vanish,
and or the compact directions are decompacted, 
the resultant amplitude consistently reproduces the 
well-known expressions available in the literature.

We examined the creation rate under 
some special cases.
We found that for the large D-string separation 
in the compact subspace and also 
the presence of the higher spacetime dimensions,
the open strings become light. Hence, 
in such scenarios, the probability
of the production of these strings increases. 
In contrast, the large D-string separation 
in the non-compact subspace leads to the 
heavy open strings, with the low creation rate.
Besides, we saw that the production rate 
in the compact spacetime may be larger than 
that in the non-compact spacetime. Occurrence 
of this fact depends on the values of the 
parameters which should satisfy the inequality 
\eqref{17}. In a special limit, the decay of the system 
in the non-compact spacetime is negligible.

%%%%%%%%%%%%%%%%%%%%%%%%%%%%%%%%%%%%%
%%%%%%%%%%%%%%%%%%%%%%%%%%%%%%%%%%%%%
\appendix
\section{Boundary state derivation and its remarks}
\label{app}
\numberwithin{equation}{section}
\setcounter{equation}{0}
%%%%%%%%%%%%%%%%%%%%%%%%%%%%%%%%%%%%%
%%%%%%%%%%%%%%%%%%%%%%%%%%%%%%%%%%%%%

This appendix is devoted to the construction of the
boundary state that describes a dressed, unstable D1-brane
with a transverse rotation in the presence of a constant
Kalb-Ramond background.
The formulation is based on the following 
$\sigma$-model action,
\begin{eqnarray}
\label{A1}
S = - \frac{1}{4\pi \alpha'}\Bigg\{ \int_\Sigma
{\rm d}^{2}\sigma \left(\sqrt{-h}h^{ab}g_{\mu\nu}
+ \epsilon^{ab} B_{\mu\nu} \right) \partial_a
X^{\mu}\partial_b X^{\nu}
- \int_{\partial\Sigma}
{\rm d}\sigma \left( \boldsymbol F
+ \boldsymbol{U} \right)\Bigg\}.
\end{eqnarray}
Here, $\Sigma$ denotes the worldsheet
of the closed string with the
metric $h_{ab} = \eta_{ab} = \text{diag}(-1, 1)$.
The boundary of the worldsheet is
displayed by $\partial \Sigma$.
The closed string is assumed to
originate from a static D$p$-brane.
The spacetime coordinates are partitioned such that
$\{x^\alpha \,|\, \alpha = 0, 1\}$ show the directions
that are parallel to the worldsheet of the D-string, while
$\{x^i \,|\, i = 2, \cdots, d-1\}$
correspond to the transverse directions.
The action includes surface terms
$\boldsymbol{F} = F_{\alpha \beta}
X^\alpha \partial_\sigma X^\beta$ and
$\boldsymbol{U} = U_{\alpha \beta} X^\alpha X^\beta$,
in which $F_{\alpha \beta}$
represents the constant field strength of the $U(1)$ gauge
potential, and the tachyon matrix
$U_{\alpha \beta}$ is symmetric and constant.

The variation of the action yields the 
following equations for
the boundary state, associated with the foregoing setup
\begin{eqnarray}
&~& \left( {\partial }_{\tau }X^{\alpha}
+\mathcal{F}^\alpha_{\;\;\;\beta}\partial_\sigma X^\beta
+ U^\alpha_{\;\;\;\beta} X^{\beta }\right)_{\tau =0}
|\mathbb B \rangle=0 ,
\nonumber\\
&~& (X^i -y^i)_{\tau =0}|\mathbb B \rangle=0 , \label{A2}
\end{eqnarray}
where $y^i$ denotes the position of the D-string,
in which we shall put $y^2=0$, and
$\mathcal{F}_{\alpha \beta} 
= F_{\alpha \beta}-B_{\alpha \beta}$
is the total field strength.

Now we impose a transverse spatial rotation
to the D-string. Thus, we define $x^2$ as the
horizontal axis and $x^1$ as the vertical axis. 
At $t = 0$, the
coordinate $x^1$ is along the D-string,
while $x^2$ is perpendicular to it.
The D-string then undergoes a
counterclockwise rotation with a constant
angular velocity $\omega$
around an axis normal to the plane $x^1x^2$.
A coordinate system $\{x'^\mu\}$ is affixed to the D-string,
such that at any given time, the
planes $x^1 x^2$ and $x'^1 x'^2$ have the same origin and
remain perfectly coincident.
This setup leads to the following transformations between
the original and rotating coordinates
\begin{eqnarray} \label{A3}
&~& x'^2 = x^2 \cos(\omega t)
+x^1 \sin (\omega t),
\nonumber\\
&~& x'^1=-x^2 \sin (\omega t)
+x^1\cos(\omega t),
\nonumber\\
&~& x'^{\bar i}=x^{\bar i}, \qquad
{\bar i} \in \{3,\cdots, d-1\}.
\end{eqnarray}

It is crucial to emphasize that
the arguments of the {\it cosine} and
{\it sine} functions is $\omega t$
but not $\omega X^0$. Here, the rotation is defined via 
the time ``$t$'', which 
is the eigenvalue of the zero-mode operator $x^0$. 
This ``$t$'' has no dependence on 
worldsheet coordinates $\sigma$ or 
$\tau$, ensuring the equations of motion 
and also the boundary state equations hold for both 
string coordinates $X^\mu$ and $X'^\mu$. 
Consequently, if the angular velocity 
is wrongly redefined as
$\omega=\frac{{\rm d}\theta}{{\rm d}X^0(\sigma, \tau)}$,
we receive a position dependent 
angular velocity $\omega(\sigma, \tau)$.
In this case, each point along the emitted closed string
from the rotating D-string exhibits
a distinct angular velocity. This
violates the assumption of a uniform angular
velocity for the rotation of the D-string.

Now for the tachyon matrix $U_{2 \times 2}$ we assume
$U_{01} = U_{11} = 0$ and $U_{00} \neq 0 \equiv U'$.
By plugging Eq. \eqref{A3} into Eq. \eqref{A2},
along with the general solution of the equation of
motion for $X^\mu (\sigma, \tau)$,
we receive the boundary state
equations in terms of the zero modes
and oscillators of the closed string coordinates.
For the zero-mode part, we obtain
\begin{eqnarray}
\label{A4}
    && \left(\alpha' p^0 - \mathcal E L^1 \cos \omega t
    + \mathcal E L^2 \sin \omega t \cos \omega t
    - \frac{1}{2}U' x^0\right)|\mathbb B\rangle^{(0)}_{(t)}
    = 0 , \\
\label{A5}
    && \left(p^1 \cos\omega t - p^2 \sin \omega t \right)|
    \mathbb B\rangle^{(0)}_{(t)} = 0,\\
\label{A6}
&& \left(L^1\sin \omega t + L^2\cos \omega t \right)
|\mathbb B\rangle^{(0)}_{(t)} = 0,\\
\label{A7}
    && \left(x^2\cos \omega t + x^1\sin \omega t \right)
|\mathbb B\rangle^{(0)}_{(t)} = 0,\\
\label{A8}
    && \left(x^{\bar i}
    - y^{\bar i}\right)|\mathbb B\rangle^{(0)}_{(t)}=0,\\
\label{A9}
&& L^{\bar i}|\mathbb B\rangle^{(0)}_{(t)}=0.
\end{eqnarray}
In the non-compact directions $L^\mu$ is zero. However,
for the compactified directions there is
$L^\mu = \alpha'(p^{\mu}_L - p^{\mu}_R)=(NR)^\mu$,
and the corresponding momentum is given by
$p^\mu = (MR^{-1})^\mu$. Here, $N^\mu$ and $M^\mu$
stand for the winding number and
momentum number of the emitted closed string state. Besides,
if the direction $x^\mu$ is compact, $R^\mu$
represents its radius of the compactification.

Some other comments are in order: From the Eq. \eqref{A5},  
we obtain $p^1 = p^2 = 0$. 
In the same way, if the directions $x^1$ and $x^2$
are non-compact, Eq. \eqref{A6} tells us that 
$L^1$ and $L^2$ identically vanish. While if
these directions are compact we receive 
$N^1 = N^2 =0$. In this case the closed strings cannot wind 
around these directions. If the direction $x^{\bar i}$ 
is non-compact, Eq. \eqref{A9} 
becomes trivial, i.e., $L^{\bar i}$ identically is zero. 
Besides, if this direction  
is compact, Eq. \eqref{A9} gives $N^{\bar i}=0$, i.e., 
closed strings cannot wrap around the 
$x^{\bar i}$-direction \footnote{We should note that the 
compactification radii are the fixed parameters of the 
background spacetime geometry. 
They are invariant under the rigid 
rotation of the D-branes, 
which merely changes the brane's orientation 
within the background.}.
Moreover, from Eq.~\eqref{A4} we obtain the relation 
$\alpha' p^0 = \mathcal{E} \frac{1}{2} U' x^0$. 
Presence of the nonzero $p^0 $
suggests an effective potential, 
influenced by the background tachyon field.

The solution of the Eqs. \eqref{A4}-\eqref{A9}
can be collectively written as
\begin{eqnarray}
\label{A10}
|\mathbb B \rangle^{(0)}_{(t)} &=& \frac{1}{\sqrt{U'}}
\int_{-\infty}^{+ \infty} {\rm d}p^0
\exp\left[\frac{i\alpha'}{U'} (p^0)^2 \right]
\;\delta(x^2 \cos\omega t + x^1 \sin\omega t)
\nonumber\\
&\times & |p^0 \rangle \otimes| p^1 =0 \rangle
\otimes | p^2 =0\rangle\otimes
\prod_{\bar i} \delta (x^{\bar i}
- y^{\bar i}) | p^{\bar i}_L = p^{\bar i}_R = 0\rangle,
\end{eqnarray}
where the prefactor $1/\sqrt{U'}$ comes from the disk
partition function \cite{36}. The exponential term in
$|\mathbb B \rangle^{(0)}_{(t)}$
demonstrates the influence of the tachyon field.

For oscillatory portion,
the boundary state equations find the feature
\begin{eqnarray}
&& \left[ \left(1-i \frac{U'}{2m}\right) \alpha_m^0
+ \left(1+ i \frac{U'}{2m}\right) \tilde\alpha_{-m}^0
+ \left(\beta_m -\tilde \beta_{-m}\right)
\mathcal E \cos \omega t \right]
| \mathbb B\rangle^{\rm osc}_{(t)} = 0,
\nonumber\\
&& \left[\left(\alpha_m^0 - \tilde \alpha^0_{-m}\right)
\mathcal E \cos \omega t + \beta_m+ \tilde\beta_{-m}
\right] | \mathbb B\rangle^{\rm osc}_{(t)} = 0,
\nonumber\\
&& \left(\gamma_m - \tilde\gamma_{-m}\right)
| \mathbb B\rangle^{\rm osc}_{(t)} = 0,
\nonumber\\
&& \left(\alpha^{\bar i}_m - 
\tilde\alpha_{-m}^{\bar i}\right)
| \mathbb B\rangle^{\rm osc}_{(t)} = 0,
\end{eqnarray}
where $m \in \mathbb Z -\{0\}$ and
$\mathcal E \equiv F_{01} - B_{01}$. Besides
the transformed oscillators are defined as
\begin{eqnarray}
\beta_m = \alpha^{1}_m\cos\omega t
-\alpha^{2}_m\sin\omega t ,
\nonumber\\
\tilde\beta_m = \tilde\alpha^{1}_m\cos\omega t
-\tilde\alpha^{2}_m\sin\omega t,
\nonumber\\
\gamma_m = \alpha^{2}_m\cos \omega t
+ \alpha^{1}_m\sin\omega t ,
\nonumber\\
\tilde\gamma_m = \tilde\alpha^{2}_m\cos \omega t
+ \tilde\alpha^{1}_m\sin\omega t .
\end{eqnarray}
These oscillators possess the 
following commutation relations
\begin{eqnarray}
[\beta_m , \beta_n]=[{\tilde \beta}_m , 
{\tilde \beta}_n]=
[\gamma_m , \gamma_n]=[{\tilde \gamma}_m , 
{\tilde \gamma}_n]
= m \delta_{m,-n},
\end{eqnarray}
and the other commutators among them are zero.

By applying the conventional method,
the oscillating part of the boundary 
state finds the solution
\begin{eqnarray}
| \mathbb B\rangle^{\rm osc}_{(t)} &=&
\frac{T_1}{2}
\left(\prod_{n=1}^\infty \Phi_{n}^{-1}(t)\right)
\exp\Bigg\{ - \sum_{m=1}^\infty \frac{1}{m} \Big[
\mathbf{A}_{(m)} \alpha_{-m}^0 \tilde\alpha_{-m}^0
- \alpha^{\bar i}_{-m} \tilde \alpha^{\bar i}_{-m}
- \gamma_{-m} \tilde\gamma_{-m}
\nonumber\\
& + & \mathbf{B}_{(m)} \beta_{-m} \tilde\beta_{-m}
+ \mathbf{C}_{(m)} \left( \alpha_{-m}^0 \tilde\beta_{-m}
- \beta_{-m} \tilde\alpha_{-m}^0 \right) \Big] \Bigg\}
|0\rangle \otimes \widetilde{|0\rangle},
\label{A14}
\end{eqnarray}
The mode-dependent terms, which are
induced by the tachyon matrix, are defined as
\begin{eqnarray}
&& \Phi_m = 1+ i (2m)^{-1} U' -\mathcal E \cos \omega t,
\nonumber\\
&& \mathbf{A}_{(m)} = -\kappa^{-1}_{(m)} \chi_{(m)},
\nonumber\\
&& \mathbf{B}_{(m)}
= 1+ 2 \kappa^{-1}_{(m)} \mathcal E^2 \cos^2\omega t,
\nonumber\\
&& \mathbf{C}_{(m)} = 2 \kappa^{-1}_{(m)}
\mathcal E \cos \omega t,
\nonumber\\
&& \kappa_{(m)} = 1 - i (2m)^{-1} U'
+ \mathcal E^2 \cos^2\omega t,
\nonumber\\
&& \chi_{(m)} = 1 + i (2m)^{-1} U'
- \mathcal E^2 \cos^2\omega t.
\end{eqnarray}

It is important to note that the total
boundary state, associated with the D-string,
consists of the zero-mode part \eqref{A10},
the oscillatory part \eqref{A14},
and also the contribution
of the conformal $(b,c)$-ghosts. The latter component
is explicitly independent of both the
background fields and rotational dynamics of the D-string.

%%%%%%%%%%%%%%%%%%%%%%%%%%%%%%%%%%%%
%%%%%%%%%%%%%%%%%%%%%%%%%%%%%%%%%%%%
%   THE BIBLIOGRAPHY     %%%%%%%%%%%
%%%%%%%%%%%%%%%%%%%%%%%%%%%%%%%%%%%%
%%%%%%%%%%%%%%%%%%%%%%%%%%%%%%%%%%%%

\end{document}